# The Influence of a Shock on the Nucleosynthesis Developing during the Explosion of a Low-Mass Neutron Star

© 2024 г.    I. V. Panov[1*], A. Yu. Ignatovskiy[1,2**], A. V. Yudin[1]

[1] «Kurchatov Institute» National Research Center, Moscow, 123182 Russia
[2] Moscow Institute of Physics and Technology (National Research University), Dolgoprudnyi, Moscow oblast, 141701 Russia

## ABSTRACT

The pattern of nucleosynthesis during the explosion of a low-mass neutron star formed in a close binary system in the stripping scenario is considered. In the scenario considered the shock arising during the explosion is shown to strongly heat the expanding neutron star matter. The heavy nuclei produced at the preceding stage of nucleosynthesis are partially destroyed as a result of a sharp increase in the role of photonuclear reactions. It is shown that even short-term heating of the matter by the shock can exert a noticeable influence on the results of the synthesis of elements in the r-process in the inner crust matter, while explosive nucleosynthesis gives rise to new elements in the outer crust matter with mass numbers A from 50 to 130.

---

[*] E-mail: igor.panov@itep.ru
[**] E-mail: lirts@phystech.edu

## 1. INTRODUCTION

After the historical joint identification of the gravitational wave signal GW170817 and the gamma-ray burst GRB170817A (Tanvir et al. 2017) as well as the observational confirmation of the production of actinides in this event (Watson et al. 2019), the merger of neutron stars (NSs) is believed to be one of the main cosmic scenarios in which the r-process is realized (Cowan et al. 2021).

Therefore, studying the details of the evolution of a close binary system at its final stages has become even more important for evaluating the results of nucleosynthesis in such scenarios.

The general scheme, i.e., the approach of NSs that ends with their contact and merger, has not undergone any noticeable changes after its first implementation (Devies et al. 1994). However, the interaction of two NSs in a close binary system may also proceed differently. The stripping scenario (Clark and Eardley 1977), which differs from the merger scenario that gained acceptance, was historically the first to be proposed. It consists in the process of relatively slow mass transfer from one component to the other and ends with the explosion of the low-mass component that has become unstable. This scenario, which had remained in oblivion for many long years, has turned out to be surprisingly successful in describing the parameters of the peculiar gamma-ray burst GRB170817A (see, e.g., Blinnikov et al. 2021, 2022).

In the classical merger scenario the formation of a massive hot NS is accompanied by the ejection of cooling, highly neutronized matter in the form of jets or a wind. In contrast, in the stripping scenario a low-mass NS (LMNS) is formed through the mass transfer from the less massive companion to the more massive one, which loses its stability and explodes as its mass decreases below the critical one. As a result of the explosion, all of the LMNS matter with a mass $\sim 0.1 M_\odot$ is ejected into the surrounding medium, enriching it with heavy elements (Panov and Yudin 2020). Recall that in the NS merger (NSM) scenario the mass of the ejecta is $\sim 10^3 - 10^4 M_\odot$ (Rosswog et al. 1999), while the bulk of the neutron-rich matter remains "locked" in the massive NS (or black hole) formed at the location of the close binary system.

Based on a numerical implementation of the stripping scenario (Yudin 2022), it was shown that all heavy elements could be produced when a LMNS is disrupted (Panov and Yudin 2020, 2023), while their yield depends on different model parameters (Yudin et al. 2023; Ignatovskiy et al. 2023). In this case, the final abundance of elements looks slightly differently than in the NSM scenario (Freiburghaus et al. 1999) predominantly due to different nucleosynthesis dynamics as the matter expands during the explosion. A consideration of the delayed explosion of a LMNS after the loss of part of the mass by it (Yip et al. 2023) confirms the results of our current and previous (Panov and Yudin 2020) nucleosynthesis calculations in the stripping scenario.

In this paper we consider the nucleosynthesis that proceeds in the crust of a LMNS as it expands during its explosion. The r-process begins in the expanding, fairly cold, highly neutronized matter (Panov and Yudin 2020, 2023; Yudin et al. 2023). However, after some time, this matter is heated by the arrived shock to

$T > 2 \times 10^9$ K, and the r-process is interrupted. As a result of a sharp increase in the temperature, the production of heavy elements acquires a different pattern for a time $\sim 10$ ms, changing noticeably the isotopic composition of the crust matter, with some of the heaviest nuclei being dissociated. After the shock passage, the temperature of the expanding matter drops, and the r-process continues in the presence of necessary conditions (the number of electrons per baryon is $Y_e < 0.4$).

This paper consists of several sections. In Section 2 we briefly describe the LMNS explosion hydrodynamics and the corresponding parameters that define the initial conditions and the r-process kinetics. In Section 3 we describe the nucleosynthesis model and the characteristic features of the expansion trajectories of the matter in which the r-process develops. In Section 4 we consider the quantitative results of nucleosynthesis and discuss how it proceeds on different LMNS matter expansion trajectories.

## 2. THE LMNS FORMATION AND EXPLOSION SCENARIO

The final evolutionary stages of a system of NSs have long been considered both as a source of short gamma-ray bursts and as a place for the r-process (Lattimer and Schramm 1974; Blinnikov et al. 1984; Becerra et al. 2023). However, in almost all of the performed multidimensional hydrodynamic simulations the masses of the NSs were close and fairly large, $M > M_\odot$, and the result of their interaction was their coalescence into one object, i.e., a merger (Korobkin et al. 2012; Rosswog et al. 2014; Martin et al. 2015). Indeed, the radius of such NSs depends weakly on the mass (Lattimer and Prakash 2001), and on contact they behave like two droplets of liquid, merging into one object – a supermassive NS or a black hole.

However, if the system of NSs is fairly asymmetric (Kramarev and Yudin 2023), i.e., the masses of the components differ significantly, then the stripping scenario can be realized (Clark and Eardley 1977). When the system's components approach each other, the lower-mass NS is the first to overfill its Roche lobe and begins to overflow to the more massive companion. During such a mass transfer it can reach the lower NS mass limit ($\sim 0.1 M_\odot$; see, e.g., Haensel et al. 2007) and explode to produce a gamma-ray burst (Blinnikov et al. 1984, 1990). The explosive disruption of a LMNS was simulated in a number of works (Blinnikov et al. 1990; Sumiyoshi et al. 1989; Colpi et al. 1989). In this paper we rely on the calculations (Yudin 2022) in which the equations of relativistic hydrodynamics (Hwang and Noh 2016) were used to solve the explosion problem. The gravitational field was assumed to be weak enough for the general relativity effects to be unimportant, but the matter velocities and the energy density were not assumed to be low. To simulate the LMNS explosion, the original equations were transformed to a form suggesting the spherical symmetry of the problem and the Lagrangian form for the convenience of their numerical solution.

The structure of the crust of the formed LMNS before its explosion is presented in Fig. 1a. The logarithm of the density $\log \rho$ is plotted against the radius r (the corresponding mass coordinates m are given on the upper axis). Part of the NS core and its inner and outer crusts are shown. The dots with numbers from 1 to 15 mark the initial positions of the trajectories along which our nucleosynthesis calculations were performed (see Table 1).

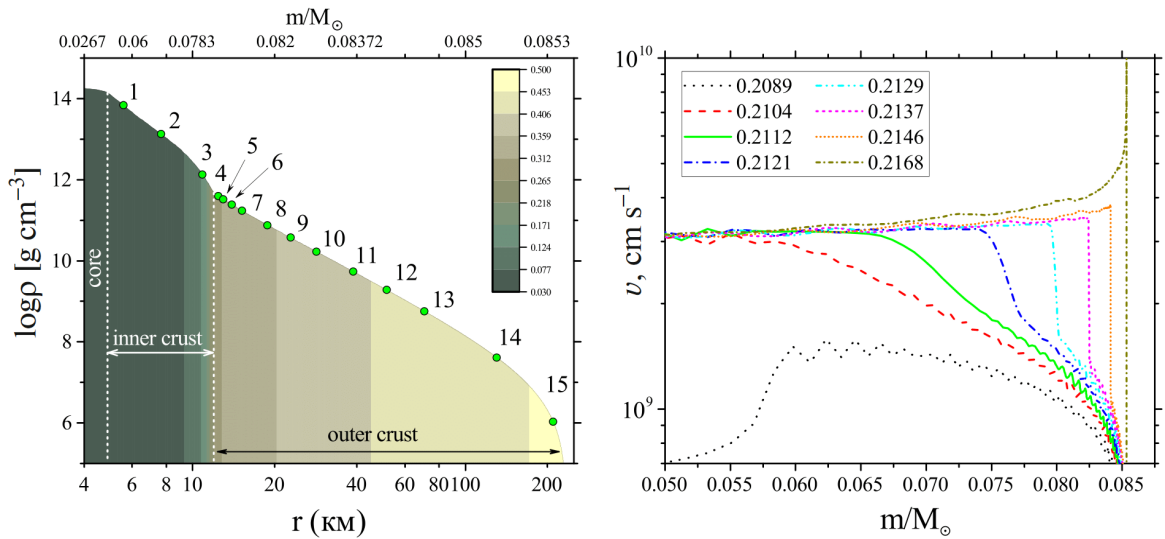

**Figure 1:** (a) The structure of a LMNS before its explosion – logarithm of the matter density $\log \rho$ versus NS radius $r$. The color panel corresponds to $Y_e$ in the matter. (b) Velocity $v$ versus mass coordinate m for several instants of time during the explosion. The values of the time cuts are given in seconds. For details, see the text.

In Fig. 1b the matter velocity v is shown as a function of the mass coordinate m for several instants of time t (the values of t are shown in the inset) measured from the time at which the star lost its hydrodynamic stability. The process of shock generation, propagation, and cumulation during its breakout is clearly seen. To estimate the conditions along each trajectory considered, it is useful to compare the left and right panels in Fig. 1 using the Lagrangian coordinate m to establish the correspondence.

After the explosion, the NS matter expands with a different velocity under the action of the shock (see Fig. 1), with the shock passage leading to short-term heating of the medium to temperatures $T \sim 10^{10}$ K or more at the maximum (see Table 1). Because of the high expansion velocity, the temperature of the shock-heated neutron-rich matter in the NS crust drops rapidly. The influence of such heating absent in the NSM model on the kinetics of nucleosynthesis reactions will be considered below. The initial characteristics of the NS crust matter for different trajectories are given in Table 1. The temporal evolution of the chemical composition for the most typical of them will be considered in detail below.

The initial characteristics of the NS crust matter for different trajectories are given in Table 1. The temporal evolution of the chemical composition for the most typical of them will be considered in detail below.

So far we did not consider the nucleosynthesis in the expanding NS core matter due to the difficulty of describing the decompression of nuclear-density matter from the central regions and determining the seed nuclei. In addition, we calculated the nucleosyn- thesis off-line on the already constructed trajectories from hydrodynamic simulations (Yudin 2022). Such a practice is a universally accepted one in calculations within the NSM model. The nearly spherically symmetric formulation of the problem of a LMNS explosion in the stripping model gives hope that self- consistent joint simulations of hydrodynamics and nucleosynthesis can be implemented. We are going to perform such simulations in the foreseeable future. Their result will probably be additional heating of the NS matter from nonequilibrium beta decays, but the exact magnitude of the effect is difficult to estimate in advance.

To calculate the initial LMNS composition, we used the BSk25 equation of state (Pearson et al. 2018). This equation of state based on the Skyrme functional complies with all of the present-day observational constraints. The dependence of the results of nucleosynthesis simulations on the equation of state being used was discussed in detail previously (Ignatovskiy et al. 2023).

Using the approximation of cold catalyzed matter (see, e.g., Haensel et al. 2007) leads to the fact that the initial composition of the matter on all trajectories (see Table 1) consists of an isotope of one element (naturally, different for different trajectories), while the structure of the NS crust consists of mononuclear layers (see, e.g., Fig. 1 in Yudin et al. (2023)). In reality, of course, even for isolated NSs the nuclear composition of the matter can slightly differ from the equilibrium one (see, e.g., Potekhin and Chabrier 2021). This is especially true for a LMNS: first, because of the tidal matter heating and, second, due to the fact that the LMNS matter experiences decompression, even though relatively slow (with a characteristic time scale $\sim 1$ s), in the stripping process. Here the composition of the matter must be determined based on the solution of the corresponding kinetic equations. So far we leave this very interesting problem for a future study.

The subnuclear-density inner crust matter (trajectories 1-3) is composed of heavy, highly neutron-rich exotic nuclei surrounded by free neutrons. These nuclei decay during the explosion and matter decompression. To calculate the nucleosynthesis in this region, we used one of the possible models for the decay of such nuclei (Yudin et al. 2023) in which the dominant decay channel is the evaporation of neutrons leading to a decrease in the atomic number $A$ with the conservation of the charge number $Z$. In this case, a nucleus with the maximum possible number $A = A_{max}(Z)$ for the chosen mass model was produced. The composition of the seed nuclei was determined in this way, while the evaporated neutrons increased the density of free neutrons. At the same time, the local value of Ye in the matter remained unchanged. For the influence of the decay model of exotic nuclei in the inner crust on the results of nucleosynthesis, see Yudin et al. (2023).

During the nucleosynthesis along trajectories 1 and 2 the inner crust matter is heated by the shock weakly, and a full r-process develops in them, while the matter expanding along trajectory 3 and the outer crust trajectories is heated by the shock strongly as the LMNS is disrupted. When the threshold temperatures for charged particles are exceeded, the ther- monuclear reactions with nucleons and alpha particles become decisive in the production of new elements. We will consider the influence of the temperature on the nucleosynthesis on these trajectories in more detail.

**Table 1:** Parameters of the evolution trajectories for the inner (1–3) and outer (4–15) NS crusts. The values of $T^{max}$ correspond to the maximum temperatures in all time for each of the trajectories, $\log \rho^{max}$ is the maximum logarithm of the density, $Y_e$ is the number of electrons per baryon, $R$ is the coordinate of the trajectory center measured from the NS center, $\Delta M$ are the masses of the matter on the trajectories in fractions of the solar mass $M_\odot$.

| Trajectory | Composition | $T^{max}/10^9$ K | Lg $\rho^{max}$ | $Y_e$ | $R$, km | $\Delta M, 10^{-4} M_\odot$ |
|---|---|---|---|---|---|---|
| 1 | $^{180}$Ce+1205n | 0.10 | 13.84 | 0.042 | 5.50 | 170.85 |
| 2 | $^{152}$Sn+735n | 0.81 | 13.13 | 0.056 | 7.65 | 155.17 |
| 3 | $^{152}$Sn+208n | 6.75 | 12.13 | 0.139 | 10.58 | 35.29 |
| 4 | $^{122}$Sr | 8.63 | 11.59 | 0.310 | 12.32 | 2.64 |
| 5 | $^{120}$Sr | 10.01 | 11.52 | 0.312 | 13.09 | 2.79 |
| 6 | $^{122}$Zr | 10.57 | 11.39 | 0.327 | 13.92 | 2.45 |
| 7 | $^{124}$Mo | 11.44 | 11.24 | 0.339 | 15.46 | 5.33 |
| 8 | $^{78}$Ni | 13.76 | 10.88 | 0.359 | 18.52 | 6.76 |
| 9 | $^{80}$Zn | 14.74 | 10.58 | 0.375 | 22.49 | 5.06 |
| 10 | $^{82}$Ge | 15.54 | 9.10 | 0.390 | 28.05 | 6.25 |
| 11 | $^{84}$Se | 16.83 | 8.94 | 0.405 | 38.15 | 7.03 |
| 12 | $^{86}$Kr | 17.94 | 8.73 | 0.419 | 50.75 | 3.98 |
| 13 | $^{64}$Ni | 17.81 | 8.50 | 0.438 | 71.00 | 5.22 |
| 14 | $^{62}$Ni | 16.32 | 7.92 | 0.452 | 127.74 | 4.61 |
| 15 | $^{56}$Fe | 10.66 | 6.56 | 0.464 | 202.33 | 0.36 |

## 3. THE NUCLEOSYNTHESIS MODEL AND SHOCK PARAMETERS

For our numerical simulations of the r-process along the evolution trajectories characterized by time-dependent parameters, including the density and the temper-

4ature, we used the kinetic network previously implemented in the SYNTHEZ code (Nadyozhin et al. 1998). It allows the number densities of all the nuclei involved in the nucleosynthesis to be determined. In the later version of the nucleosynthesis model implemented in the SYNTHER (nucleoSYNThesis of HEavy elements in the R-process) code (Korneev and Panov 2011) the fission reactions were supplemented by a more proper allowance for the mass distribution of fission product nuclei and allowance for their return to the r-process as new seed nuclei, leading to the establishment of a quasi-steady flow of nuclei under certain conditions (Panov 2016).

In real scenarios the conditions typical for the r-process can be realized, in particular, after explosive nucleosynthesis during the collapse of supernovae (Woosley et al. 1994) or in other explosive processes, for example, during the explosion of a LMNS (Yudin 2022; Ignatovskiy et al. 2023). In this case, seed nuclei for the r-process from the iron-peak elements to zirconium are produced. Therefore, the capabilities of the code were enhanced, and a number of additional reactions typical for explosive nucleosynthesis at high temperatures were included in the model (Arnett 2000; Rauscher and Thielemann 2000). The list of weak-interaction reactions was expanded, and the interactions of nucleons and nuclei with electrons (Panov et al. (2016) important at high temperatures and densities were taken into account in the nucleosynthesis model. The region of the nuclei for which the reactions with charged particles were taken into account was also expanded, and the database of reaction rates based on temperature-independent approximations of these quantities (Rauscher and Thielemann 2000) was updated.

Since the reaction rates of the listed processes, which determine the eigenvalues of the Jacobi matrix for the system of differential equations implemented in our nucleosynthesis codes, differ in absolute value by many orders of magnitude, the system of nucleosynthesis equations is a classical example of a stiff system of ordinary differential equations. For its numerical integration we used the Gear (1971) method, with the previously developed matrix inversion software package (Blinnikov and Dunina-Barkovskaya 1994) having been used for its implementation. As the main algorithm we used the predictor–corrector method with an automatic choice of the step and the accuracy order of the method that was implemented in both codes used by us with an internal check for the conservation of the number of nucleons and charge.

We determined the boundaries of the region of the nuclides involved in the nucleosynthesis: $Z_{min} = 1$, $Z_{max} = 110$, $A_{min}(Z)$, and $A_{max}(Z)$, according to the mass model being used: the extended Thomas–Fermi plus Strutinsky integral method (Aboussir et al. 1995) or the finite-range droplet model (FRDM) (Moeller et al. 1995). Depending on the model, we determined the boundaries of the region under consideration and the total number of nuclei $N$ involved in the nucleosynthesis, which varied from 5800 to 6300.

The nuclear reaction rates, which are the coefficients in the differential equations, were calculated using the same mass models. The list of nuclear reactions includes all pair reactions with neutrons, protons, alpha particles, and gamma-ray photons; beta decay and beta-delayed processes, such as the emission of several neutrons during

– 7 –

beta decay and delayed fission; induced and spontaneous fission; and a number of other important reactions, such as the $3-\alpha$ reaction and the $^{12}C$, $^{16}O$, and $^{28}Si$ burning reactions.

The developed scheme allows one to effectively calculate the nucleosynthesis in various scenarios at temperatures $T < 7 \times 10^9$ K and densities $\rho < 10^{12}$ g cm$^{-3}$. Our main nucleosynthesis calculations were made using the known beta decay, delayed neutron emission, and alpha decay rates (Moeller et al. 1997, 2003) as well as the thermonuclear reaction rates (Rauscher and Thielemann 2000). The fission was taken into account for all nuclei of chemical elements with $Z > 82$, although only the actinides have significant fission rates. The capture rates of neutrons by heavy nuclei (for elements with $Z > 82$) and the neutron-induced fission rates are based on the calculations of the reaction cross sections in the Hauser–Feshbach model (Panov et al. 2010), as are the delayed fission rates (Panov et al. 2005). The spontaneous fission was taken into account according to phenomenological models (Panov et al. 2013).

In the most popular NSM scenario, while expanding and cooling, the hot matter of the jets ejected during the formation of a massive compact object creates conditions typical for the development of the r-process that continues either until the exhaustion of neutrons or until the drop in the matter density below the critical one. In the stripping scenario being considered by us the r-process begins in the matter of the exploded, fairly cool NS after the decompression of subnuclear-density matter, which after some time transforms into explosive nucleosynthesis as a result of the heating of the medium by the shock. Although the duration of the explosive nucleosynthesis is short, the intermediate composition of the chemical elements changes through photonuclear reactions. The r-process can be continued after the drop in temperature as a result of the ongoing matter expansion if the conditions necessary for it to proceed are preserved. Such conditions after the shock passage are preserved only for trajectory 3, on which the r-process with the seed nuclei produced through the explosive nucleosynthesis continues after the drop in temperature. The various regimes of nucleosynthesis and the dependence of the composition of the synthesized elements on the envelope parameters in the stripping scenario will be considered in detail in the next section.

## 4. THE PRODUCTION OF HEAVY ELEMENTS IN THE NS CRUST

Based on the stripping model, we calculated the abundances Y (A) for fifteen trajectories starting from different layers of the LMNS crust (see Fig. 1 and Table 1).

In our calculations we used an improved code database in which the limited library of thermonuclear reaction rates (Thielemann 1987) was replaced by a more complete one (Rauscher and Thielemann 2000), the different libraries were jointed more carefully, and the updated rates of neutron reactions for elements with $Z > 82$ (Panov et al. 2010) and weak interactions (Langanke and Martinez-Pinedo 2000) were used.

The stripping scenario differs from the merger scenario, in particular, by the expansion dynamics and different time dependences of the density and, most dramatically, the temperature due to the shock action. The lower matter expansion velocity at the initial stage here leads to a slower decrease in the density (Fig. 2a), while the initially low temperature increases abruptly for a short time due to the heating by the shock (Fig. 2b). Thus, the stripping scenario also differs from the NSM scenario by the type and pattern of nucleosynthesis.

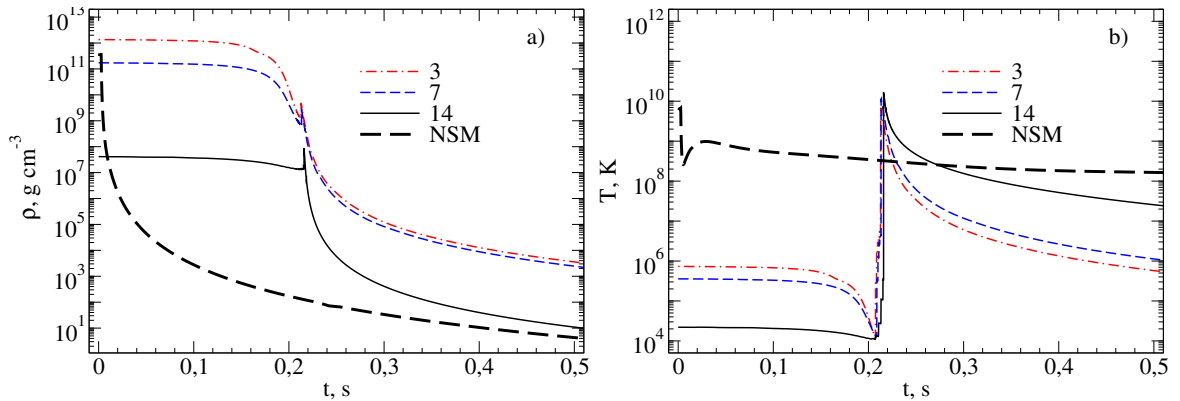

**Figure 2:** Density (a) and temperature (b) versus time $t$ in seconds in the expanding matter for different trajectories: 3, 7, and 14. To compare the matter expansion dynamics in the NSM scenario (Freiburghaus et al. 1999), we present the temperature and density evolution curves during the merger of NSs with approximately equal masses. The code of these curves is NSM.

Figure 2 shows the time dependences of the matter density and temperature for several trajectories of the outer and inner crusts. In the nucleosynthesis along these trajectories heavy elements are produced most intensively both in the r-process (Panov and Yudin 2020, 2023; Yudin et al. 2023) and in the mixed type of nucleosynthesis partly considered by Ignatovskiy et al. (2023). It can be seen from this figure that within $t_{sw}$ after the explosion on some inner-crust trajectories and some of the outer-crust trajectories in which the r-process began after the decompression of matter from a subnuclear density, the shock catches up with the expanding neutron-rich matter and heats it to temperatures at which the r-process ceases and explosive nucleosynthesis begins.

The time dependence of the density of free neutrons (Fig. 3) illustrates the nucleosynthesis dynamics when the outer envelope of the inner crust (Fig. 3a) and the envelopes of the outer crust (Fig. 3b) are ejected and, together with Fig. 2, shows the presence of medium parameters needed to maintain the r-process ($\rho > 1$ g cm$^{-3}$, $T < 2 \times 10^9$ K, $N_n > 10^{22}$ cm$^{-3}$) or their absence.

A strong r-process develops in the layers of the inner crust, but the nucleosynthesis development dynamics for these layers is slightly different. As the crust matter expands along trajectory 3, a sharp increase in $N_n$ (approximately by an order of magnitude) during its heating by the shock causes a high temperature to be maintained for a time $\Delta t \sim 0.001$ s, which is barely noticeable in Fig. 3 and insignificant relative to the preceding value of $N_n \sim 10^{33}$ cm$^{-3}$. The temperature jump interrupts the r-process, while the started explosive nucleosynthesis leads to the photodissociation of the heaviest nuclei. As the temperature decreases behind the shock, the nucleosynthesis again returns to the r-process starting from a new distribution of atomic nuclei with considerably smaller $Z$ and $A$. As a result, the blast wave along trajectory 3 (Fig. 3a) only suspends the r-process, which is resumed after the end of explosive nucleosynthesis. The new composition for the r-process starting for the second time consists of less neutron-rich nuclei, while the density of free neutrons is much lower than that before the shock arrival. Therefore, the r-process will end more rapidly than in the model with the absence of heating caused by the shock.

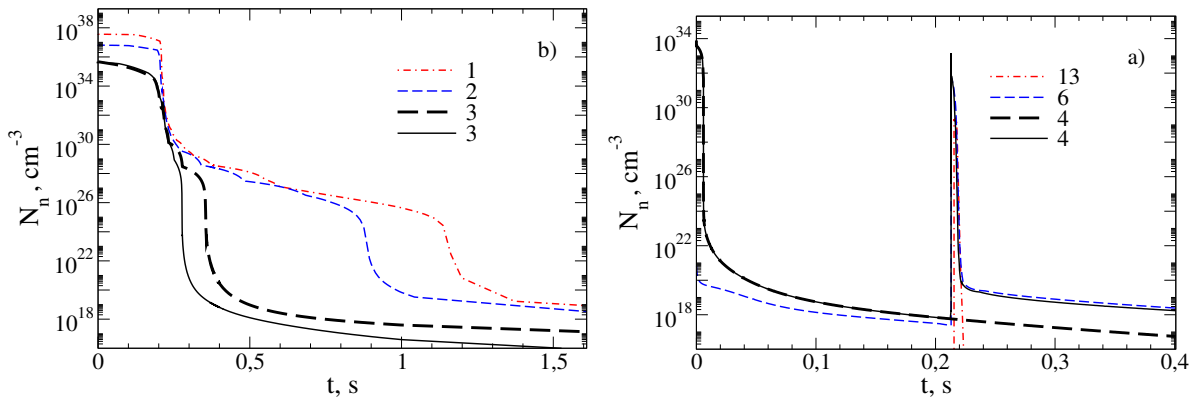

**Figure 3:** Number density of free neutrons $N_n$ in cm$^{-3}$ versus time $t$ in seconds for typical trajectories of the inner (a) and outer (b) crusts. The code of the curves is the trajectory number. The thick dashed line corresponds to the calculations without including the heating by the shock.

Much of the inner-crust matter (trajectories 1 and 2) is heated by the shock insignificantly ($T < 10^9$ K), and a strong r-process develops in them owing to the high fraction of free neutrons with the production of elements with a mass number greater than 1000 typical for it. Since the rise in temperature for these trajectories is insignificant (see Table 1), no explosive nucleosynthesis develops and no nuclei with $A < 100$ are produced.

When the outer crust is ejected, the heating of the matter by the shock leads to the creation of condi- tions for explosive nucleosynthesis (Fig. 4). Along several trajectories, with numbers 4–7, the initial high neutron number density is kept for a very short time, and even a weak r-process has no time to be realized, since the conditions for the r-process are maintained for a time shorter than the characteristic beta decay time $\tau_\beta$. Here only a weak n-process is possible (Blake and Schramm 1976), because for trajectories 4–7 the neutron capture rates (see also Fig. 3b, $0.02 < t < 0.2$) are comparable to the beta decay rates. Within the time before the shock arrival, $t_{sw}$, a distribution of nuclei that consists of the isotopes of the original nucleus and elements adjacent to it, from molybdenum to cadmium, is formed due to this process (curves 1 in Figs. 4a and 4b). In contrast, with the shock arrival explosive nucleosynthesis begins.

After the shock passage and the end of explosive nucleosynthesis, the continuation of a weak n-process is possible, since the density of free neutrons is approximately an order of magnitude higher than that before the shock arrival (see Fig. 3b). The evolution of the composition in such an n-process can be seen in Fig. 4b, when the strong even–odd effect after the shock passage (curves 3) is smoothed out (curves 4) and chemical elements up to the second peak are produced.

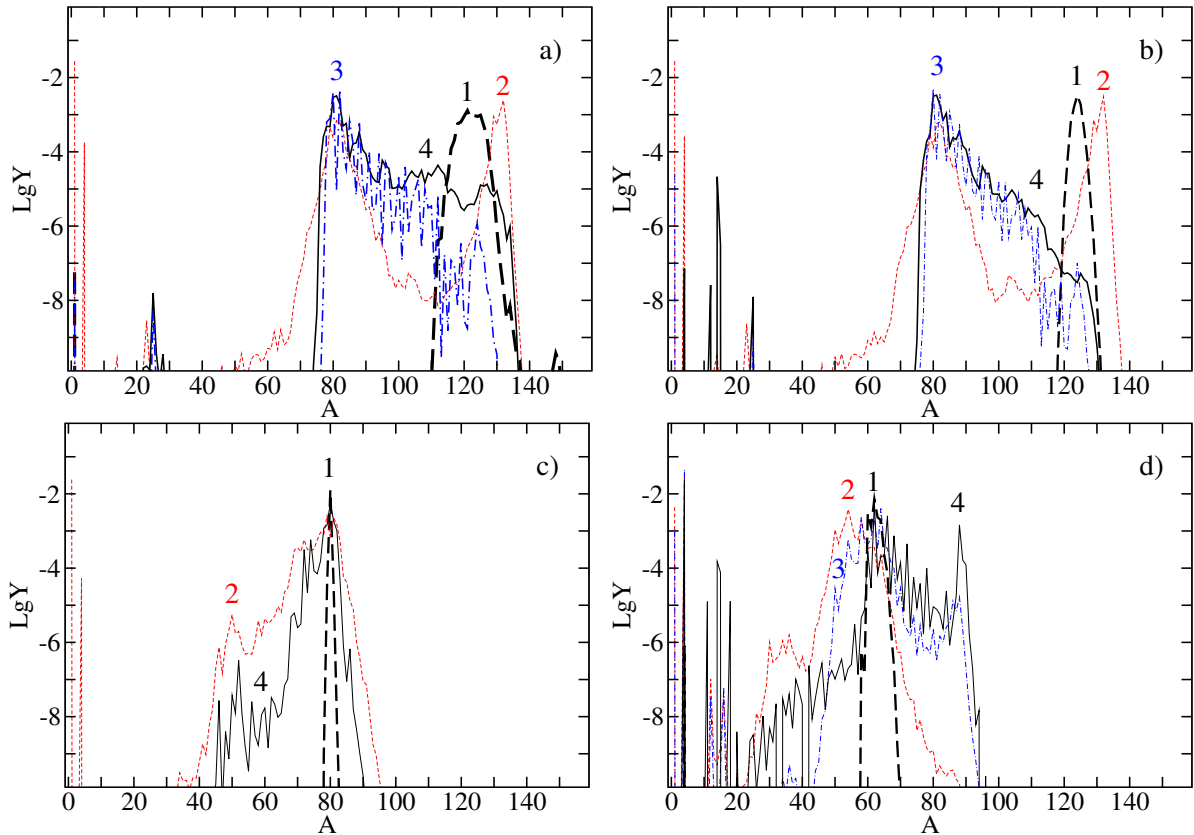

**Figure 4:** Chemical element abundance per baryon $Y = X/A$ versus atomic mass number $A$ during the shock passage along trajectories 4 (a), 7 (b), 9 (c), and 14 (d) of the outer crust: (1) before the shock arrival; (2) at the shock peak; (3) $\Delta t \approx 3 \times 10^{-3}$ s after the shock, $T \sim 2 \times 10^9$ K; and (4) the final distribution. $X$ is the mass fraction.

Note that in explosive nucleosynthesis for many of the outer-crust trajectories the amount of chemical elements increases, but their ordinal number and mass number decrease.

Note that on a short time scale during the nucleosynthesis along trajectories 4–7, apart from the intensive formation of a new abundance peak of chemical elements near $A \sim 80$, a peak in the region $(Z, A) \sim (50, 130)$ temporarily arises because of the shell effects, which is rapidly destroyed to produce new lighter elements with $A < 90$. In the long run, after the drop in temperature below $T < 3 \times 10^9$ K, heavier elements are produced in the inner part of the outer crust during the subsequent expansion of the matter in a weak n-process (curves 4 in Figs. 4a and 4b).

Figures 4c and 4d show how the isotopic composition of the matter expanding along trajectories 9–14, on which the matter is heated by the shock to higher temperatures ($T > 10^{10}$ K) than those during the expansion of the matter along trajectories 4–7 evolves after the LMNS explosion. The initial mononuclear composition of the crust along these trajectories transforms in the nucleosynthesis process insignificantly, with the production of a small amount of new chemical elements, since the n-process does not proceed due to the deficiency of free neutrons. For trajectory 9 (Fig. 4c) the initial composition is close to the equilibrium distribution and, therefore, changes little during the heating by the shock.

For the matter in which $Y_e > 0.35$ before the explosion no short-lived neutron-rich isotopes are produced, while new isotopes are produced in the region of even lighter nuclei with $A < 80$ through photonuclear reactions, which is typical for trajectories 10–15 and is illustrated for the evolution of the matter composition along trajectory 14 (Fig. 4d). Note that for this trajectory the results slightly differ from the explosive nucleosynthesis along other trajectories of the outer crust (10–15). For trajectory 14 a noticeable amount of the heavier isotope $^{88}$Sr is produced through the $\alpha$ process because of the shell effects.

Let us now consider the production of heavy elements as the inner-crust matter expands (Fig. 5). During the nucleosynthesis along trajectory 3, after the shock passage and the heating of the matter to a temperature $T \sim 2 \times 10^9$ K or higher, the r-process ceases and is replaced for a short time by explosive nucleosynthesis. The nucleosynthesis region is shifted from the line of neutron stability toward stable nuclei, changing noticeably the abundances of the isotopes synthesized before the shock passage through photoreactions and producing some number of nuclei with $A < 80$. During the subsequent expansion of the matter and its cooling the reactions with charged particles freeze and the r-process is resumed. The change in abundance as a result of such a delay is shown in Fig. 5a. It can also be seen from this figure that in the absence of heating (dashed curve) no light and intermediate elements are produced.

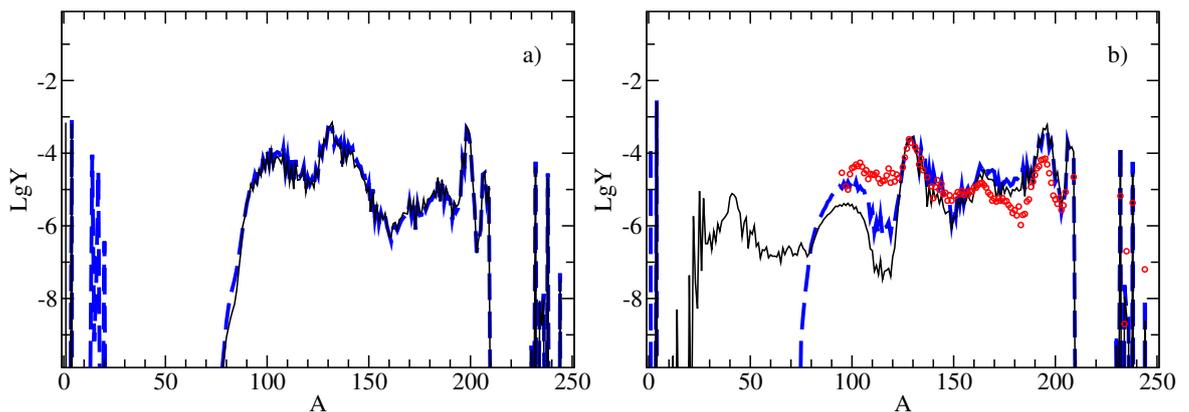

**Figure 5:** Calculated chemical elements abundance per baryon $Y = X/A$, versus atomic mass number $A$ in the inner crust: (a) along trajectory 1 (solid curve); (b) along trajectory 2 (dashed curve); along trajectory 3 with (solid curve) and without (dashed curve) heating. The dots mark the solar elemental abundances in relative units. $X$ is the mass fraction.

Along trajectories 1 and 2 the temperature rises insignificantly during the shock passage, and the nucleosynthesis proceeds only through the r-process. A strong feedback due to fission can be seen from the shape of the curve, and complete burning of nuclei lighter than the cadmium-peak elements is observed. Some number of light nuclei (mostly carbon and oxygen) is produced only during the evolution of the chemical composition through the $(\gamma,\alpha)$ and $3\alpha-$ reactions along trajectory 2, during the heating by the shock to temperatures $T \sim 10^9$ K. Note that the sum of all the calculated mass fractions $X_i = AY_i$ in Figs. 4 and 5 is equal to 1.

## 5. CONCLUSIONS

Previously, we (Panov and Yudin 2020, 2021) pointed out that the shock briefly heats the matter of the NS envelopes to temperatures of $10^{10}$ K or more, depending on the equation of state being used in the stripping model (Ignatovskiy et al. 2023), but the details of this heating and its influence on the nucleosynthesis were not discussed in detail.

In this paper we studied the pattern of nucleosynthesis during a LMNS explosion in the stripping scenario for a number of important evolution trajectories of the matter expanding from different zones of the crust. The initial compositions of the trajectories differed greatly in chemical elements (see Table 1), neutron excess (the initial value of $Y_e$ changed from 0.042 for the initial composition of trajectory 1 (inner crust) to 0.452 for trajectory 15 (outer crust)), and heating (to $T \sim 1.8 \times 10^{10}$ K) for trajectories 12 and 13. It turned out that such a wide set of parameters leads to three different types of nucleosynthesis attributable to short, but strong heating of the matter by the shock.

(1) The r-process in the inner crust is character- ized by a significant role of fission. As the dense neu- tronized matter is decompressed along trajectories 1– 3, the synthesis of heavy elements under the action of neutrons begins. At the peak of the shock part of the inner-crust matter expanding along trajectory 3 is heated to a temperature $T \sim 6 \times 10^9$ K. The rates of the reactions with charged particles increase by orders of magnitude, and the r-process transforms into explosive nucleosynthesis.

Although the duration of such nucleosynthesis is short ($\sim 10$ ms), the composition of the produced nuclei changes, especially for isotopes with $A < 120$, compared to the calculation without heating (see Fig. 5), with a noticeable amount of lighter chemical elements being produced. After the ongoing expansion of the envelope and the cooling of its matter, the r-process is resumed. The role of heating is actually reduced to a delay in the propagation of the nucleosynthesis wave, effectively reducing the duration of the r-process (trajectory 3) and leading to better agreement with observations in the region of the rare-earth peak ($A \sim 160-180$) compared to the scenarios without intermediate heating of the matter during the r-process.

For two trajectories originating deep in the inner crust (1 and 2) the heating of the matter by the shock is insignificant, and the nucleosynthesis through the r-process, as during the nucleosynthesis along trajectory 3, leads to the burning of initial seed nuclei with the establishment of a quasi-equilibrium flow of nuclei between the fission region and the region of fission product nuclei. Such cycling of the nucleosynthesis along all inner-crust trajectories crust leads to the production of a comparable amount of heavy chemical elements with $A > 90$.

(2) At the beginning of the explosion, because of the small ratio of free neutrons to seed nuclei, the density $N_n$ during the nucleosynthesis along trajectories 4–7 of the outer crust drops rapidly. This leads to an increase in the number of isotopes from the initial mononuclear composition through the n-process. During the shock passage and the heating of the medium the rates of the photonuclear reactions and the reactions with charged particles increase by orders of magnitude. As a result, new channels of reactions involving photonuclear reactions open up, explosive nucleosynthesis begins, the nuclei produced at the preceding neutron-rich cold stage are partially destroyed, and lighter chemical elements are produced. The n-process developing after the end of explosive nucleosynthesis in these layers of the outer crust increases the amount of heavy elements up to the region of the second peak.

(3) In the remaining layers of the outer crust explosive nucleosynthesis predominantly occurs, in which lighter elements compared to the initial mononuclear composition are produced (trajectories 8–15).

On the whole, it can be concluded that the heating of the inner (trajectory 3) and outer (trajectories 4–15) crust matter by the shock leads to a pattern of nucleosynthesis different from the nucleosynthesis through the r-process alone (trajectories 1 and 2).

Comparing two different scenarios, the stripping and NSM ones (see, e.g., Martin et al. (2015) and references therein), it is worth noting that both in the NSM jets and in the expanding matter of the inner crust during the explosion of a LMNS in the stripping scenario heavy elements from fission products to actinides are produced in the r-process. In the wind from a hot massive NS (NSM) and in the outer layers of an exploding LMNS (stripping) lighter elements are produced. These scenarios do not contradict each other and are varieties of the evolution of a close binary NS system. Only further observations will help to establish the contribution of each of them to the total abundance of heavy elements.

**ACKNOWLEDGMENTS**. We are grateful to M.V. Zverev and A.I. Chugunov for the fruitful discussions of the properties of ultradense neutron matter and the NS crust. We also thank the anonymous referees for the remarks that allowed the pre- sentation of our results to be improved significantly.

**FUNDING**. This work was performed within the State assignment of the «Kurchatov Institute» National Research Center.

**CONFLICT OF INTEREST**. The authors of this work declare that they have no conflicts of interest.